\documentclass[aps,prl,amsmath,amssymb,twocolumn,superscriptaddress]{revtex4}
\usepackage{dcolumn}
\usepackage{graphicx}

\begin{document}

\title{Model for Assembly and Gelation of Four-Armed DNA Dendrimers}
\author{Francis W. Starr}
\affiliation{Department of Physics, Wesleyan University, Middletown,
  Connecticut 06459, USA}
\author{Francesco Sciortino}
\affiliation{Dipartimento di Fisica and INFM Udr and Crs:Soft: Complex Dynamics 
in Strucured Systems, Universita` di Roma La Sapienza, Piazzale Aldo Moro 2, I-00185 Rome, Italy}

\date{\today}

\begin{abstract}
We introduce and numerically study a model designed to mimic the bulk behavior of a system composed of single-stranded DNA dendrimers.  Complementarity of the base sequences of different strands results in the formation of strong cooperative intermolecular links.  We find that in an extremely narrow temperature range the system forms a large-scale, low-density disordered network via a thermo-reversible gel transition.  By controlling the strand length, the  gel transition temperature can be made arbitrarily close to the percolation transition, in contrast with recent model systems of physical gelation.  This study helps the understanding of self-assembly in this class of new biomaterials and provides an excellent bridge between physical and chemical gels. 
\end{abstract}

\maketitle

The ``bottom-up'' construction of new materials is one of the central aims of nanotechnology~\cite{glotzer}. The recent synthesis of nanoparticle building blocks functionalized with specifically designed  oligonucleotides has opened new possibilities for the assembly of networked materials~\cite{mirkin,milam,kiang}.  In this approach,  a number of single strands of  DNA are grafted on the surface of
micron-sized particles.  The interactions between the particles are controlled by the addition of DNA strands in solution which have complementary sequences to the DNA strands grafted on the particles.  The novelty of the approach arises from the possibility of using the high sensitivity and selectivity  of complementary strand recognition to tune on and off the interparticle bonding.   The use of  DNA  sequences to establish interparticle interactions provides an optimal choice for the construction of three-dimensional supramolecular assemblies, since DNA strands can self-assemble into long and fairly rigid helices based on sequence complementarity~\cite{seeman}.   New materials can be designed by modulating the length of the binding sequence, the length of the grafted single strand, or the number of grafted strands.  DNA-decorated colloids potentially exhibit extremely rich behavior, since, in addition to modifying base sequencing, the colloidal properties also can be altered.  This tremendous number of possible choices makes DNA-linked assemblies
one of the most versatile and promising new soft-matter materials and thus calls
for theoretical~\cite{dema,tracenko} and numerical studies of these systems.

One of the key aims of such studies is the prediction of the three-dimensional
self-assembled structure of these networked materials. 
It has been considered disappointing that most DNA-decorated colloidal dispersions form highly disordered aggregates, instead of 
crystal-like structures~\cite{kiang,biancaniello}.  These new materials are prone to form, in a reversible way, gels, i.e. disordered arrested states at low densities.   
In contrast to chemical gels, whose understanding has progressed much further due to the conceptual simplicity introduced by the infinite bond lifetime and the theoretical framework of percolation theory~\cite{stauffer,torquato},  the thermo-reversible formation of colloidal gels in the absence of phase separation and crystallization is still an open problem in soft condensed matter~\cite{rubinstein,jack}.  Recent studies have suggested that the generation of a thermo-reversible physical gel at low temperature $T$ --- i.e. conditions such that the bond lifetime is longer than the observation time and stresses can effectively propagate through the sample
--- is facilitated by limiting the preferred number of bonding neighbors (valency)~\cite{tan-stock,zac05}.  Such a constraint decreases the surface tension of a cluster aggregate, thereby destabilizing the phase separation process.
Novel biomaterials in which a specific (and small) number of complementary DNA strands are attached to a common center~\cite{seeman,stewart} should naturally lead to a limited valency, and hence are an interesting group of potential gel-forming systems.  The possibility of controlling both the number of arms (the valency) and the bonding energy (via the number of complementary sequences) makes these materials among the best candidates for checking the propensity to form gels and test recent explanations of physical gel formation~\cite{zac05}, as well as for exploring the types and properties of self-assembled biomaterials.


\begin{figure}[h]
\centerline {\includegraphics[width=3.2in]{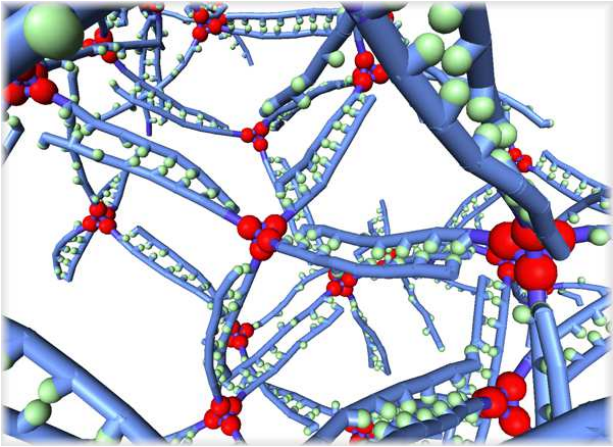}  }
\caption{Snapshot of the simulation at low $T$ where most base pairs are bonded.  The red spheres indicate the tetrahedral core of each DNA tetramer.  The blue tubes indicate the bonds along a single strand of the DNA, and the small green spheres represent attractive base-pair force sites.  Hence the regular pairing of these green spheres shows the proclivity for complementary base pairs to attract each other. \vspace{-0.5cm}}
\label{fig:pictures}
\end{figure}

In this Letter we introduce a model  designed to mimic the tetrameric DNA complexes recently synthesized and discussed in Ref.~\cite{stewart}, but
whose bulk behavior has not yet been explored.  The model is not designed to be chemically accurate, but should qualitatively reproduce the physical phenomena of DNA assembly.  Using the model, we study the bulk behavior of a
system of many tetramers and find that in an extremely narrow $T$ range, the system forms a low-density disordered network via a thermo-reversible gel transition.  In contrast with previously studied
cases of thermoreversible gelation~\cite{zac05,fscollaborators,jack2,jack}, the gel and percolation transitions of our model can be made arbitrarily close by exploiting the entropic contribution to bond formation made possible
by the complementary DNA binding.

Each molecular unit of the model is composed of a tetrahedral hub tethering four identical DNA-like strands composed of eight connected monomers;  we will refer to this molecular unit as a tetramer.  The ordering of the bases beginning from the tetrahedral core is A-C-G-T-A-C-G-T;  A, C, G, and T are the standard abbreviations for the bases of DNA.  Bases of type A bond only with type T, and type C bonds only with type G.  We choose this sequencing since it offers the possibility of forming bonds between different tetramers in which all eight sites along a strand are paired.   All pairs of sites interact via a purely repulsive potential obtained by truncating and shifting the Lennard-Jones (LJ) potential
\begin{equation}
V_{\rm sf}(r) = V_{\rm LJ}(r) - V_{\rm LJ}(r_c) - (r-r_c)
\left.\frac{dV_{\rm LJ}(r)}{dr}\right|_{r=r_c}.
\label{eq:shifted-force}
\end{equation}
The cutoff $r_c=2^{1/6} \sigma$, where $\sigma$ is LJ length parameter.    Neighboring monomers (those along a strand and in the central tetrahedral hub) are connected via a FENE anharmonic spring potential $V_{\rm FENE} = -k(R_0^2/2) \ln (1-(r/R_0)^2)$, where the bond strength $k=30$, and the maximum bond extension $R_0 = 1.5$, as used in ref.~\cite{benn1} to study coarse-grained polymers.  To model the characteristic rigidity of the DNA strands, we add a three-body
potential of the form $k_{\ell} (1-\cos\theta)$, where
$\theta$ is the angle defined by three consecutive monomers.  A value of $k_{\ell}=5$ allows for moderate flexibility of the strands, but prevents strands in the same tetramer complex from becoming entangled.

\begin{figure}[h]
\centerline {\includegraphics[width=3.4in]{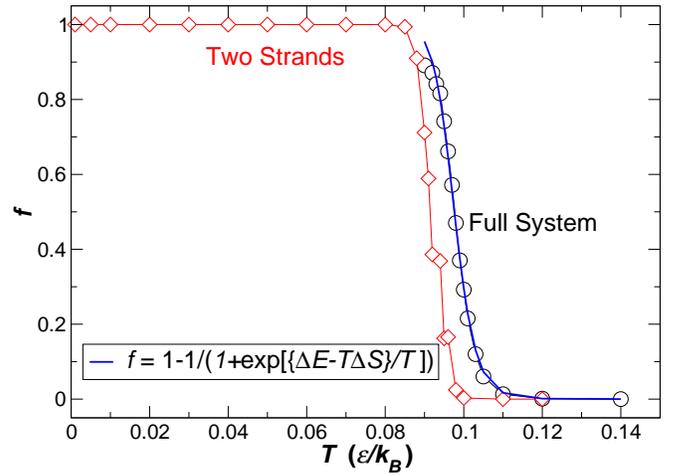}}
\caption{The fraction $f$ of bonded DNA strands as a function of $T$.  The figure shows $f$ for the full system of 200 tetramers as well as for two isolated strands.  The crossover from the unbonded to nearly fully bonded state is extremely sharp, but not discontinuous and it is well described by the two-state  relation
$f=1-1/(1+e^{(\Delta E -T \Delta S)/T)})$, with $\Delta E=3.53$ and $\Delta S = 36.1$.  We confirm absence of hysteresis on reheating.  We also performed a preliminary study on 64 tetramers which showed statistically identical behavior, suggesting there are no significant finite size effects.  Hence, the crossover appears to be largely insensitive to system size.  In order for the plots to be comparable, both systems have a strand density of 0.0056. \vspace{-0.8cm}}
\label{fig:bonds-energy}
\end{figure}

To simulate bond formation between complementary bases, each monomer along the strand has an additional ``bonding'' force site that carries the information about the base type.  Attractive interactions are included only between the bonding sites of complementary bases.   The bonding sites are connected to the monomer core using the same FENE potential that links the strands together.  The interactions between complementary bonding sites are modeled via a LJ potential as in Eq.~(\ref{eq:shifted-force}), but the truncation distance $r_c=2.5 \sigma_{aa}$ to include attractions.  We choose $\sigma_{aa} = 0.35 \sigma$ so that the bonding site is almost completely contained in the repulsive shell of the monomer core.  This choice prevents the bonding site from connecting to more than one complementary base.  Interactions between non-complementary bases are also given by Eq.~(\ref{eq:shifted-force}) with $r_{c}=2^{1/6}\sigma_{aa}$, so that interactions are purely repulsive.  Fig.~\ref{fig:pictures} shows a snapshot of the system when many base pairs are bonded.  
We present our results in reduced units where length is in units of $\sigma$, time in units of $\sigma\sqrt{m/\epsilon}$, $T$ in units $\epsilon/k_B$ ($k_B$ is Boltzmann's constant), and entropy is in units of $k_B$.

We simulate the model via molecular dynamics calculations to generate data for the configurations and velocities of the constituent particles as a function of time.  We use this information to examine thermodynamic, structural, and dynamic properties.  
We study a system of $N=200$ tetramers (a total of 13600 force sites) in a box of length $L = 52.41$, resulting in a molecular number density $n=1.39 \times 10^{-3}$.  For this density, the approximate average distance between the centers of tetramers $\ell = n^{-1/3} = 8.96$.  This separation is ideal for the formation of networks, since the distance between two bonded cores is $\approx 9$, including the 8 base pairs in the strand and the tethering monomers at the core.  Each simulation is performed at a fixed density and $T$, and we control $T$ using the Nose-Hoover method~\cite{frenkel}.  At each $T$ we simulate 2 independent systems to improve our statistics.  At the lowest $T$ studied, our simulations extend to more than $10^{7}$ time steps in order to access equilibrium behavior.  However, we point out that this lowest $T$ may still exhibit modest aging effects, but these will not affect our overall conclusions.  To accelerate the overall speed of the simulations, we use a 3-cycle velocity Verlet version of rRESPA multiple time step algorithm with the forces separated into rapidly varying bonded forces and more slowly varying non-bonded forces~\cite{frenkel}.  The time step for the bonded forces is 0.002.

\begin{figure}[t]
\centerline {\includegraphics[width=3.4in]{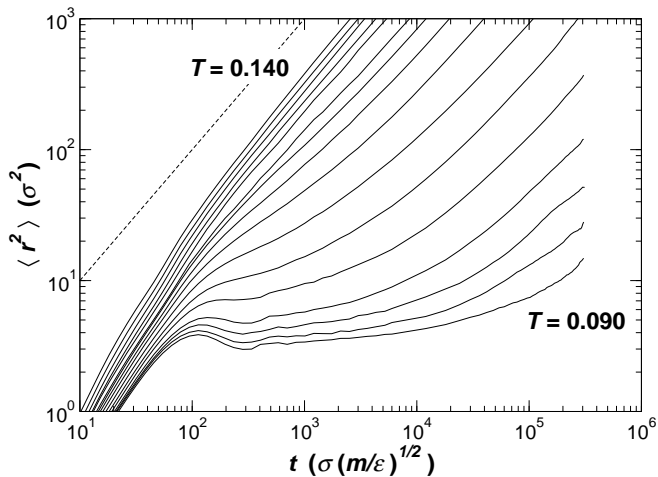}}
\caption{Mean-squared displacement $\langle r^2 \rangle$ of the tetramer center of mass.  
The dotted line indicates the asymptotic linear behavior of $\langle r^2 \rangle$ that is expected when tetramers are diffusive.  The first 4 decades in time are not shown, since the dynamics at this time scale are trivial.  From top to bottom, the $T$ we study are $T=0.140$, 0.120, 0.110, 0.105, 0.103, 0.101, 0.100, 0.099, 0.098, 0.097, 0.096, 0.095, 0.094, 0.093, 0.092, 0.090. \vspace{-0.6cm}}
\label{fig:msd}
\end{figure}

To explore gel formation in this system, we study several different $T$ values. Fig.~\ref{fig:bonds-energy} shows the fraction $f$ of bonded strands as a function of $T$.   Since the attractive well between complementary base pairs is quite narrow, we say that a base pair is bonded if the energy between the pair is negative;  we consider two strands to be bonded if at least half of the complementary base pairs of two strands are bonded.  Fig.~\ref{fig:bonds-energy} demonstrates that the range of the transition to a highly bonded state occurs over a narrow range $0.09 \lesssim T \lesssim 0.11$ --- only 2\% of the energy of a single bond. A simple  two-state
model, in which the open and bonded states of the strand are attributed to an  entropy change $\Delta S = 36$ and a energy change $\Delta E = 3.5$, accurately describes the $T$-dependence of $f$.  The large $\Delta S$ value shows that, in this model, bonding of strands freezes between four and five entropy units per base. This entropic contribution is responsible of the nearly first-order nature of the crossover to the bonded state, a feature not observed in  recently studied physical gel models~\cite{zac05,jack2,jack}, the consequences of which will become apparent shortly.  For comparison, Fig.~\ref{fig:bonds-energy} also shows $f(T)$ for two isolated strands.
While in the bulk system the competition between bonding arms prevents reaching the fully bonded state,  the fully bonded ground state is easily reached at low $T$ for the case of two isolated strands.  
 
\begin{figure}[t]
\centerline {\includegraphics[width=3.4in]{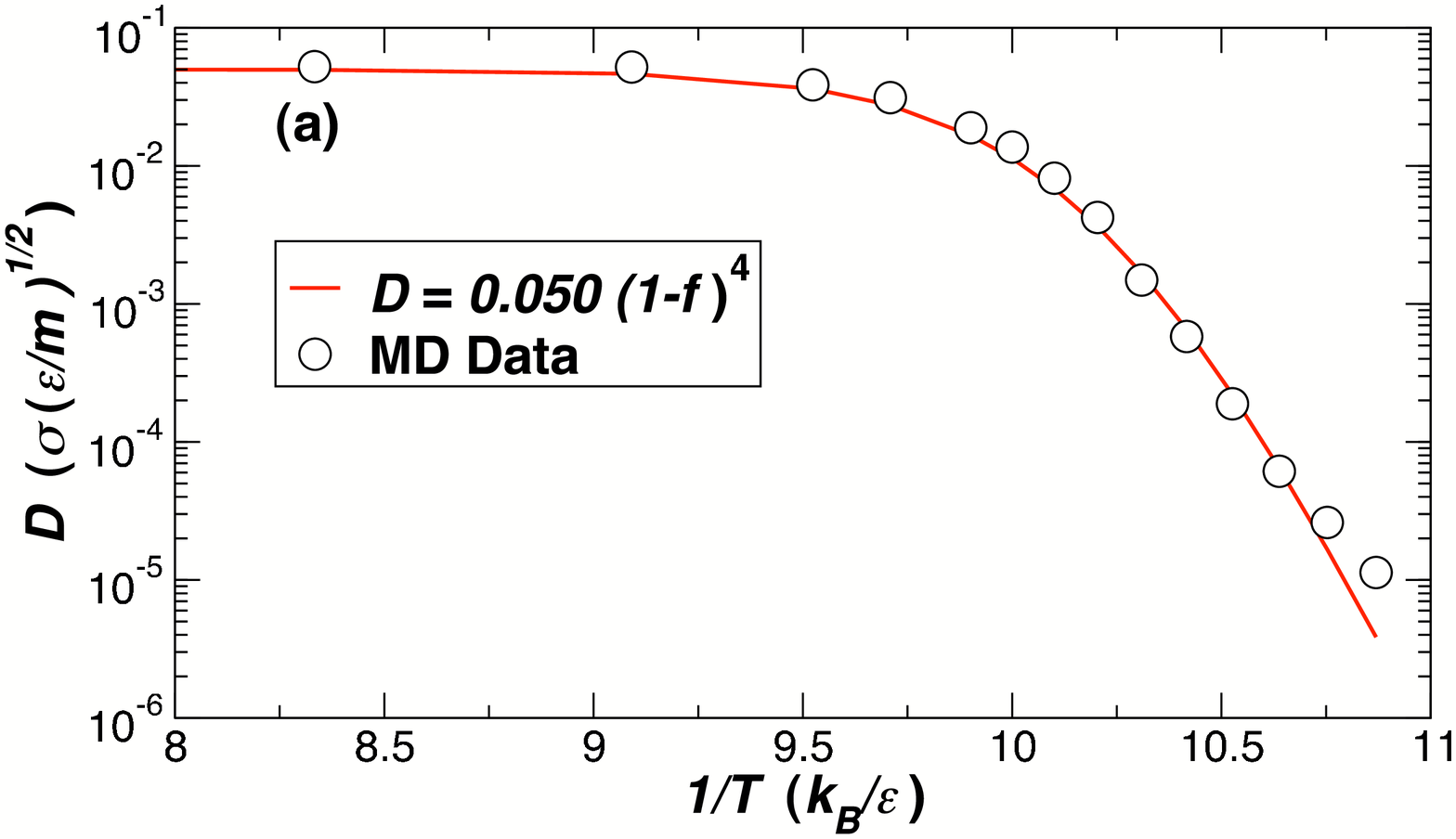}}
\centerline {\includegraphics[width=3.4in]{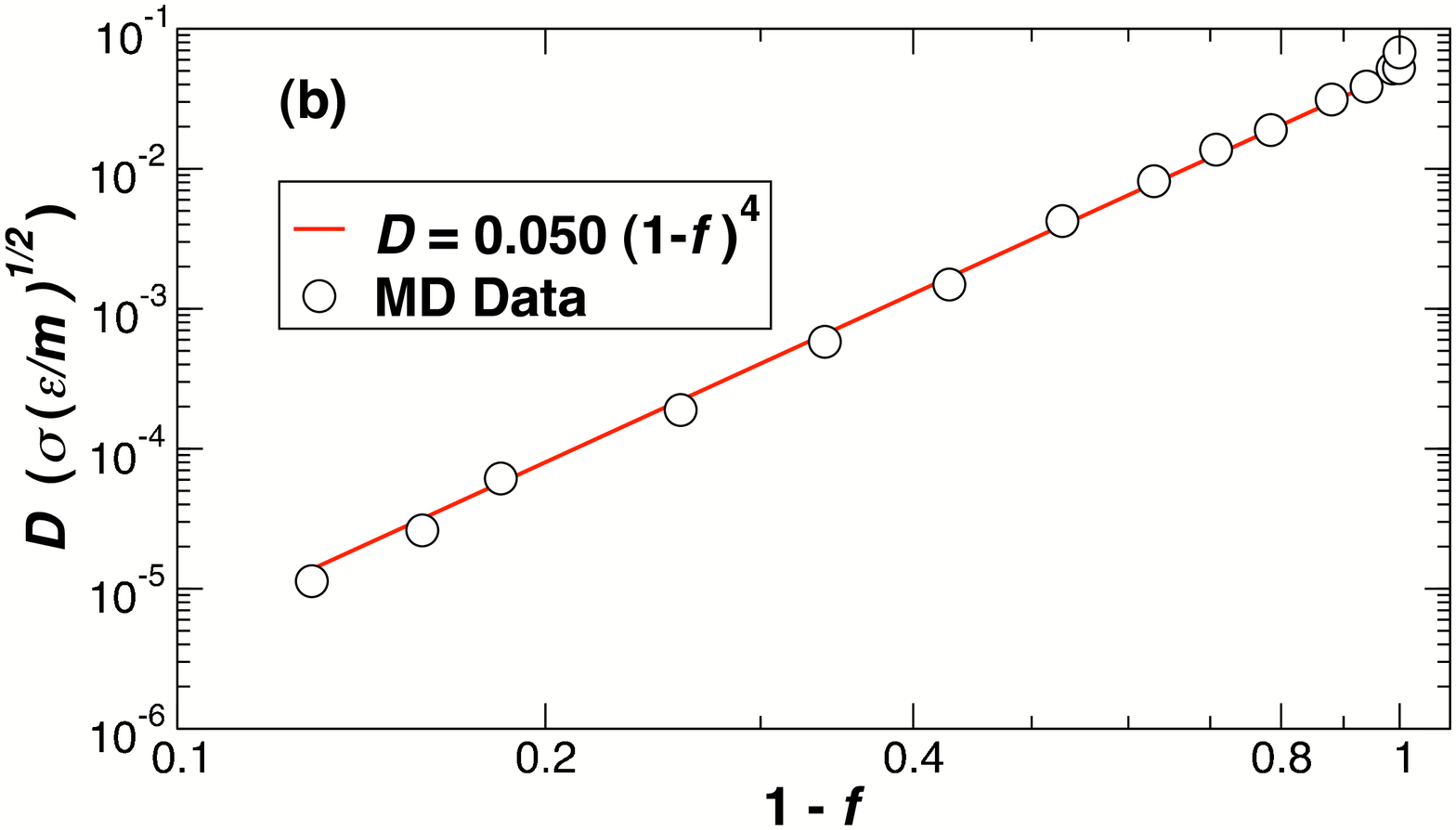}}
\caption{The relation of diffusion constant $D$ to $T$ and bonding fraction. (a) Arrhenius plot showing that $D$ follows an Arrhenius law at low $T$.  The line uses the 2-state fitting form from fig.~\ref{fig:bonds-energy}. (b) Dependence of $D$ on $f$; the line is the best fit between $D$ and the raw data for $1-f$, without using the 2-state model.  We find a power law relation  $D = D_{0} (1-f)^{4}$ over all $T$, where $D_{0} = 0.050$.  Hence $D$ is controlled by the fraction of unbonded tetramers.  \vspace{-0.8cm}}
\label{fig:dynamics}
\end{figure}

We next quantify the ``freezing'' of the dynamic properties expected to occur once the gel state has formed.   The mean-squared displacement $\langle r^2 \rangle$ of the tetramer center of mass (Fig.~\ref{fig:msd}) demonstrates that at high $T$, where $f\approx 0$, tetramers are able to diffuse with little hindrance.  Over this narrow $T$ range,  $\langle r^2 \rangle$ becomes strongly hindered.  
Fig.~\ref{fig:dynamics}(a) shows the diffusion constant $D$ calculated from the asymptotic behavior of $\langle r^2 \rangle = 6Dt$   as a function of $1/T$.  When few bonds are present, the $T$ dependence of $D$ is very weak. The slowing of the dynamics is intimately connected to the formation of bonds in the system; in the same narrow $T$ region where bonding becomes significant, $D$ dramatically decreases.  The nearly linear behavior $D$ at low $T$ in Fig.~\ref{fig:dynamics}(a) indicates a limiting Arrhenius behavior, i.e.~$\ln D \sim 1/T$.  More importantly, we find a linear relation between $D$ and $(1-f(T))^{4}$ (Fig.~\ref{fig:dynamics}(b)).  Since $f$ is the fraction of bonded strands, $f$ also equals the probability that a strand is bonded.  Thus $1-f$ is the probability that a strand is unbonded, and so $(1-f)^{4}$ is the probability that all 4 strands of a tetramer are unbonded.   The quality of the comparison between $D$ and $(1-f)^{4}$ demonstrates that the variation of $D$ at fixed density is controlled entirely by the fraction of fully unbound tetramers. This finding mirrors the behavior recently found in a limited-valence model for colloidal gels~\cite{fscollaborators}.  Since $f(T)$ can be effectively described  by the two-state model shown in Fig.~\ref{fig:bonds-energy}, at low $T$ 
\begin{equation}
D = D_0 [1+\exp ((\Delta E - T \Delta S)/T)]^{-4} \approx D_0 e^{4 \Delta S} e^{-4 \Delta E/ T},
\label{eq:arr-explained}
\end{equation}
which explains the observed low $T$ Arrhenius behavior.

The low $T$ Arrhenius dependence of $D$ means that motion will technically only cease in the limit $T\rightarrow 0$, but  for practical purposes, the dynamics in the $T$ window studied have already become much slower than any reasonable computational time. We call the $T$ at which $D$ can no longer be estimated the gel temperature $T_{\rm gel} = 0.092$, with $f(T_{\rm gel}) \approx 0.87$ and $D(T_{\rm gel}) \approx 1.1 \times 10^{-5}$.  A much more restrictive definition of $T_{\rm gel}$, based  on a diffusion coefficient ten orders of magnitude smaller --- i.e. $D(T_{\rm gel}) \approx 10^{-15}$ --- would only move $T_{\rm gel} \approx 0.080$, a decrease of just 13\%.

As we increase the length of strands, both $\Delta E$ and $\Delta S$ will increase, since they are proportional to  the number of bonds between the strands.  Combining this with Eq.~(\ref{eq:arr-explained}) shows that increasing $\Delta S$ will cause $D$ to decrease extremely rapidly over an even more narrow range of $T$, leading to the formation of an arrested state in a nearly discontinuous manner.  Thus the sharpness of the crossover to the gel state can be tuned simply by changing the number of bases in the strand.     

In recent model systems of physical gels~\cite{jack,jack2,zac05} geometric percolation of clusters  does not correlate with dynamic arrest, since at percolation,  clusters break and reform continuously.  Thus the $T$ dependence of $D$ is not strongly influenced by the crossing of the percolation locus.
This is in sharp contrast with chemical gels (in which bonds form irreversibly),  where gelation and percolation coincide.  In our system, the close correlation between $f$ and dynamics suggests that for sufficiently long strands the percolation of the DNA network is connected with the system's dynamic arrest, even though bonding is reversible.  We locate the percolation transition using standard algorithms to partition tetramers into clusters of size $s$, identify spanning clusters, and  evaluate the distribution of cluster sizes $n(s)$.  Approaching percolation,  $n_s \sim s^{-\tau}$, with $\tau\approx-2.2$, the theoretical value expected for random bond percolation~\cite{torquato}.  We find the percolation temperature $T_{p} \approx 0.099$.  At this $T$, molecules are still able to diffuse
but $T_{p}$ is very near to $T_{\rm gel} = 0.092$.  For a chemical gel, $T_{p} = T_{\rm gel}$ so that percolation ideas can be used to understand gel formation.   Given the close correspondence between the length of stands and the activation energy for $D$, 
DNA-linked colloidal particles with controlled functionality make it possible
to tune the distance between $T_{\rm gel}$ and $T_{p}$.  The possibility of controlling in a reversible way 
the gel transition and the on-off character of the transition makes the DNA-gels optimal biomimetic materials for  delivery and release of host components.

In summary, we have shown the tendency for specifically sequenced DNA-dendrimers to assemble into amorphous gel structures.  In doing so, we demonstrated the close connection between the fraction of bonded strands and the dynamics of this new class of materials.  As a final comment, we recall that in the nanotechnology bottom-up approach~\cite{seeman} individual components are designed to assume particular tertiary structures with the aim of self-assembly into quaternary structures.  In this respect, understanding the gel propensity of the designed nanostructure is fundamental for generating three dimensional structures with desired properties. 

We thank the NSF for support under grant number DMR-0427239; F.S. 
thanks also  Miur FIRB and MRTN-CT-2003-504712.  We thank J.F.~Douglas and L.W.~McLaughlin for valuable discussions, P.~Filetici for bringing ref.~\cite{stewart} to our attention and the B.U. Center for Polymer Studies for hosting us. 

 \end{document}